\documentclass[letterpaper]{article} 
\usepackage{aaai2026}  
\usepackage{times}  
\usepackage{helvet}  
\usepackage{courier}  
\usepackage[hyphens]{url}  
\usepackage{graphicx} 
\usepackage{natbib}  
\usepackage{caption} 
\usepackage{amsmath}
\usepackage{multirow}
\usepackage{booktabs}
\usepackage{algorithm}
\usepackage{algorithmic}

\usepackage{newfloat}
\usepackage{listings}
\DeclareCaptionStyle{ruled}{labelfont=normalfont,labelsep=colon,strut=off} 
\floatstyle{ruled}
\newfloat{listing}{tb}{lst}{}
\floatname{listing}{Listing}
\title{HuiduRep: A Robust Self-Supervised Framework for Learning Neural Representations from Extracellular Recordings}
\author {
    Feng Cao\textsuperscript{\rm 1}\thanks{These authors contributed equally. Author order follows the alphabetical order of surnames.},
    Zishuo Feng\textsuperscript{\rm 2}\textsuperscript{*}\textsuperscript{†},
    Jicong Zhang\textsuperscript{\rm 1}\thanks{Corresponding authors.},
    Wei Shi\textsuperscript{\rm 1}\textsuperscript{†},
}
\affiliations {
    \textsuperscript{\rm 1}Beihang University\\
    \textsuperscript{\rm 2}Beijing University of Posts and Telecommunications\\
    kohaku@buaa.edu.cn, akatukifzs@bupt.edu.cn \\ jicongzhang@buaa.edu.cn, shiweilab@buaa.edu.cn
}

\usepackage{bibentry}

\begin{document}

\maketitle

\begin{abstract}
Extracellular recordings are transient voltage fluctuations in the vicinity of neurons, serving as a fundamental modality in neuroscience for decoding brain activity at single-neuron resolution. Spike sorting, the process of attributing each detected spike to its corresponding neuron, is a pivotal step in brain sensing pipelines. However, it remains challenging under low signal-to-noise ratio (SNR), electrode drift and cross-session variability. In this paper, we propose \textbf{HuiduRep}, a robust self-supervised representation learning framework that extracts discriminative and generalizable features from extracellular recordings. By integrating contrastive learning with a denoising autoencoder, HuiduRep learns latent representations that are robust to noise and drift. With HuiduRep, we develop a spike sorting pipeline that clusters spike representations without ground truth labels. Experiments on hybrid and real-world datasets demonstrate that HuiduRep achieves strong robustness. Furthermore, the pipeline outperforms state-of-the-art tools such as KiloSort4 and MountainSort5. These findings demonstrate the potential of self-supervised spike representation learning as a foundational tool for robust and generalizable processing of extracellular recordings.

\end{abstract}


\section{Introduction}

Neuroscientists frequently record extracellular action potentials, known as spikes, to monitor brain activity at single-cell resolution. These spikes, the extracellular voltage deflections from individual neurons, are considered the “fingerprints” of single-cell activity. By analyzing spike trains, which are sequences of temporally ordered spike times, researchers can infer neuronal coding and dynamics with millisecond precision \cite{10.3389/fninf.2022.851024}.  

However, each electrode often captures spikes from many nearby neurons, so it is crucial to sort or cluster spikes by their source \cite{dallal2016dictionary, banga2022spike}. Spike sorting is the process of assigning each detected spike waveform to its originating neuron \cite{guzman2021extracellular}. In practice, spike sorting is treated as a clustering problem on waveform features, often following the initial steps of filtering and spike detection \cite{souza2019spike}. It is a foundational step in electrophysiology that enables single-unit analysis and studies of neuronal function \cite{REY2015106}. 

In classical spike sorting pipelines, data are first preprocessed, typically filtered and normalized. Spikes are then detected, typically via threshold crossings or template matching. Subsequently, features such as waveform principal components or wavelet coefficients are extracted. The resulting feature vectors are then clustered using methods like k-means, Gaussian Mixture Model (GMM) or density-based algorithms to identify putative single units. Early automated sorters such as KlustaKwik \cite{kadir2013highdimensionalclusteranalysismasked} often required extensive manual curation due to imperfect clustering. More recent frameworks like MountainSort \cite{chung2017fully} and KiloSort \cite{NEURIPS2023_83c637c3} have improved throughput. For instance, MountainSort introduced an automatic clustering approach with accuracy comparable to, or exceeding manual sorting. Likewise, KiloSort4 \cite{pachitariu2024spike} uses template matching and deconvolution to scale sorting to hundreds of channels with high accuracy. These tools represent the state-of-the-art in spike sorting, but they still rely on conventional clustering paradigms and presuppose stable, high-quality signals.

Despite recent advances, spike sorting remains challenging under realistic conditions. Low SNR signals make spikes difficult to detect or distinguish. Nearby neurons often produce overlapping or morphologically similar waveforms, leading to "compound" spikes that violate the assumption of one spike per neuron. Electrode drift, slow movement of neurons relative to the probe, causes spike waveforms to change over time, violating the stationarity assumption. Electrode drift has been identified as a major contributor to sorting errors, and correcting for drift substantially improves sorting performance. Spatial overlap of neurons also complicates sorting: dense, high–channel-count probes produce many overlapping electrical fields, worsening the "collision" problem. 

In practice, even the best algorithms degrade under such conditions: for example, methods without explicit drift correction such as SpyKING CIRCUS \cite{10.7554/eLife.34518} and  earlier versions of MountainSort lose accuracy when drift is large. Conventional methods also struggle with diverse waveform shapes, and cross-session variability may result in inconsistent unit identities across different recording sessions \cite{doi:10.1126/sciadv.adr4155}. Thus, robustly clustering spikes in noisy, drifting data remains a key open problem.

To address these issues, we propose HuiduRep, a self-supervised representation learning framework for extracting representations of spike waveforms for spike sorting. HuiduRep learns features that are discriminative of neuron identity while being less affected by noise and drift. Inspired by recent trends in extracellular recordings representation learning \cite{NEURIPS2023_83c637c3}, HuiduRep combines contrastive learning with a denoising autoencoder (DAE) \cite{vincent2008extracting}. As a result, HuiduRep can learn robust and informative spike representations without any manual labeling. We then cluster the learned representations using the Gaussian Mixture Model to perform spike sorting. Building upon this, we further design a complete pipeline for spike sorting. The pipeline achieves robustness to low SNR and drift, and outperforms state-of-the-art sorters such as KiloSort4 and MountainSort5 on accuracy and precision across diverse datasets.

In summary, our main contributions are as follows:
\begin{itemize}
    \item We propose HuiduRep, a novel self-supervised framework that integrates contrastive learning and DAE with physiologically inspired view augmentations for robust spike representation learning.
    \item We design a complete pipeline for spike sorting that requires no ground truth labels and supports high-density probes.
    \item We evaluate our method on datasets from distinct neural structures, demonstrating its robustness, and we show that it outperforms state-of-the-art sorters.
\end{itemize}

Code and supplementary materials are available at \url{https://github.com/IgarashiAkatuki/HuiduRep}

\begin{figure*}
    \centering
    \includegraphics[width=0.98\linewidth]{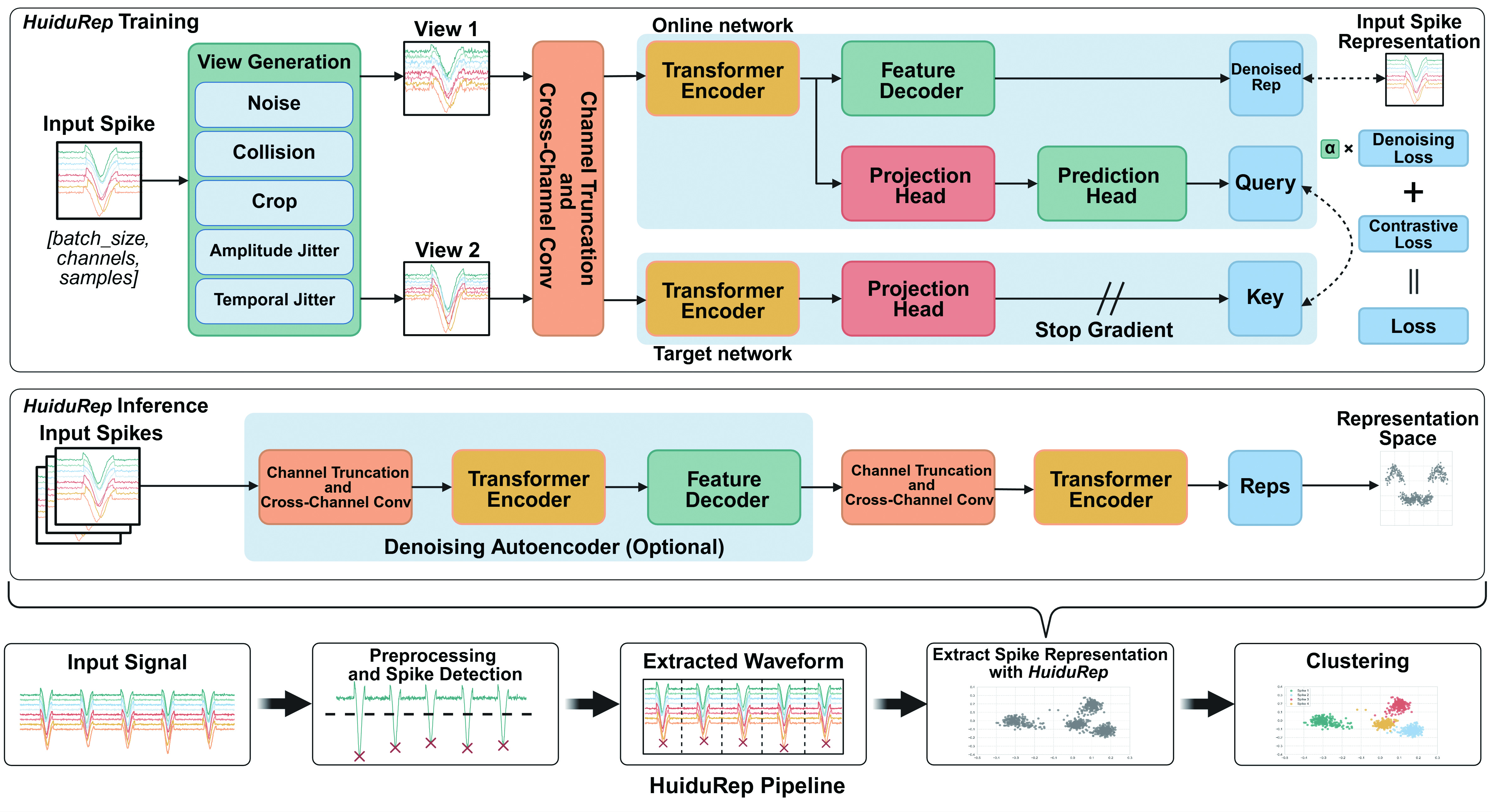}
    \caption{Overall architecture of HuiduRep and the pipeline. During training, the contrastive learning branch adapts the MoCo v3 style framework, where the \textit{query} is compared with \textit{key} and other in-batch samples (not shown in the figure due to the limited space). Only the \textit{View 1} is passed to the DAE branch for reconstruction. During inference, only the transformer encoder and DAE module are used to extract representations.}
    \label{fig:enter-label}
\end{figure*}
\section{Related Work}
\subsection{Template-based Spike Sorters} 
Template-based spike sorting algorithms remain one of the most widely used methods for processing extracellular recordings. These approaches typically detect spikes and then cluster them by matching their waveforms to a set of learned templates.

KiloSort is one of the most widely adopted template-matching sorters. It performs spike detection and sorting in a unified framework using a template matching approach combined with drift correction. Operating directly on raw data, KiloSort can handle large-scale recordings, such as those produced by high-density Neuropixels probes \cite{doi:10.1126/science.abf4588}. Its core idea is to model the recorded signals as a superposition of spatiotemporally localized templates and to iteratively infer spike times and unit identities.

Despite the success of KiloSort and other template-based methods, which often rely on handcrafted heuristics or static templates that may not generalize well to low SNRs or rare waveform variations. These limitations have motivated the development of recent deep learning-based methods, including our proposed HuiduRep, which aims to learn robust representations directly from data without relying on fixed templates.

\subsection{Representation Learning Models}
In spike sorting, effective representation of spike waveforms plays a crucial role in enabling accurate clustering, particularly in noisy and drifting recordings. Recent methods have therefore adopted representation learning frameworks to learn spike features. Among these, CEED \cite{NEURIPS2023_83c637c3} and SimSort \cite{zhang2025simsortdatadrivenframeworkspike} have emerged as two representative approaches that leverage contrastive learning to derive meaningful spike features without manual labeling.

CEED is a SimCLR-based \cite{chen2020simpleframeworkcontrastivelearning} contrastive representation learning framework for extracellular recordings. It is trained and evaluated on the IBL dataset \cite{10.7554/eLife.63711}, achieving promising performance in embedding spike features. Nevertheless, CEED is limited in this scope: it functions solely as a feature extractor and does not design a complete spike sorting pipeline. Moreover, its performance degrades sharply in embedding multiple neuron types, indicating its difficulty in capturing fine-grained inter-class distinctions.

Compared to CEED, SimSort not only proposes a representation learning model but also introduces a complete spike sorting pipeline. However, SimSort also has several limitations. For instance, it only supports 4-channel inputs, which limits its applicability to high-density probes such as Neuropixels recordings \cite{doi:10.1126/science.abf4588}. Moreover, due to its relatively small model size, the performance improvement over existing sorters remains limited, especially in noisy and drifting recording conditions.

\section{Method}
In this section, we introduce the overall architecture of HuiduRep as well as the pipeline based on HuiduRep.
\subsection{Architecture of HuiduRep}
The overall architecture of HuiduRep is illustrated in Figure 1. Inspired by BYOL \cite{grill2020bootstraplatentnewapproach} and MoCo-v3, our framework also consists of two main branches: an online network and a target network. The target network, which is frozen during training, is updated via a momentum update based on the online network’s parameters.

The key difference lies in the introduction of a DAE within the online network, which is designed to reconstruct the original spike representations from the augmented views generated by the view generation module. This DAE serves as an auxiliary module to guide representation learning. Moreover, we replace the original ResNet encoder \cite{he2015deepresiduallearningimage} with a Transformer encoder \cite{vaswani2023attentionneed}. Before feeding the input views into the encoder, we also apply cross-channel convolution to better capture the characteristics of spike waveforms. During training, only \textit{View 1} is fed into the DAE branch, while \textit{View 2} does not participate in the denoising task.

Furthermore, the contrastive learning branch adapts the MoCo-v3 style \cite{chen2021empiricalstudytrainingselfsupervised}, where representations from positive pairs (\textit{query} and \textit{key}) and in-batch negative samples are compared. For contrastive learning, we adopt the InfoNCE loss \cite{oord2019representationlearningcontrastivepredictive}, while for denoising, we employ the mean squared error (MSE) loss:
\begin{align*}
\mathcal{L}_{\rm{Contrastive}} &=-\rm{log}\frac{exp(\mathit{q\cdot k^{+}} / \tau)}{exp(\mathit{q\cdot k^{+}} / \tau) + \sum\limits_{\mathit{k^{-}}}exp(\mathit{q\cdot k^{-}} / \tau)} \\ 
\mathcal{L}_{\rm{Denoising}} &= \frac{1}{n}\sum\limits_{i=1}\limits^{n}(v_{i}-\hat v_{i})^{2}
\end{align*}
Here $q$ denotes the query vector output by the prediction head of the online network, $k^{+}$ represents the positive key generated by the target network for the same sample and $k^{-}$ refers to the negative keys, which are the outputs of other samples in the same batch passed through the target network. $\tau$ is a temperature hyper-parameter \cite{wu2018unsupervisedfeaturelearningnonparametric} for $l_{2}$ -normalized $q$ and $k$. For MSE loss, $v$ is the embedded feature obtained from the original input, while $\hat v$ is the reconstruction produced by the DAE. We apply a standard MSE loss to measure the reconstruction quality. 
The overall loss function of the model is a weighted sum of the denoising loss and the contrastive loss.
\begin{algorithm}[tb]
\caption{PyTorch Style Pseudocode of HuiduRep}
\label{alg:algorithm}
\begin{algorithmic}[0] 
\STATE \textit{\# conv: channel truncation + cross-channel convolution}
\STATE \textit{\# f\_q: encoder + projection + prediction}
\STATE \textit{\# f\_k: momentum encoder + momentum projection}
\STATE \textit{\# dae: encoder + feature decoder}
\STATE \textit{\# clf: contrastive loss function}
\STATE \textit{\# a: weight factor}
\STATE \textit{\# m: momentum coefficient}
\STATE
\STATE for x in loader:\ \ \# load data
\STATE \ \ \ \ v1,\ v2 = aug(x), aug(x)\ \ \# augmentation
\STATE \ \ \ \ v1,\ v2 = conv(v1), conv(v2)\ \ \# conv embeddings
\STATE \ \ \ \ q1,\ q2 = f\_q(v1), f\_q(v2)\ \ \# queries
\STATE \ \ \ \ k1,\ k2 = f\_k(v1), f\_k(v2)\ \ \# keys
\STATE
\STATE \ \ \ \ v = conv(x)\ \ \# conv embeddings
\STATE \ \ \ \ v\_hat = dae(v1)\ \ \# denoising branch
\STATE
\STATE \ \ \ \ loss1 = clf(q1, k2) + clf(q2, k1)\ \ \# symmetrized
\STATE \ \ \ \ loss2 = MSELoss(v, v\_hat)
\STATE \ \ \ \ loss = loss1 + a * loss2\ \ \# weighted loss
\STATE \ \ \ \ loss.backward()
\STATE
\STATE \ \ \ \ \# optimizer update
\STATE \ \ \ \ update(f\_q), update(dae), update(conv) 
\STATE \ \ \ \ f\_k = m*f\_k + (1-m)*f\_q\ \ \# momentum update
\end{algorithmic}
\end{algorithm}

Inspired by CEED, several augmentation strategies are employed to the original spike waveforms to generate input views. These include:
(1) Voltage and temporal jittering, which introduces small perturbations in both voltage amplitude and timing;
(2) Channel cropping, where a random subset of channels is selected to create partial views of the original waveforms; 
(3) Collision, where noisy spikes are overlapped onto the original waveforms to simulate spike collisions; and
(4) Noise, where temporally correlated noise is added to the waveforms to generate noised views. This Noise method is employed only for generating \textit{View 1}, enhancing the robustness and performance of the DAE. The detailed view augmentation strategy is provided in the supplementary material.

During inference, HuiduRep uses the encoder from the contrastive learning branch to extract representations of input spikes for downstream tasks. In certain cases, the DAE can be optionally applied before the encoder to further enhance the overall performance of the model.
\subsection{Spike Sorting Pipeline}
Based on HuiduRep, we propose a complete pipeline for spike sorting. As illustrated in Figure 1, our pipeline consists of the following steps:
(1) Preprocessing the raw recordings by removing bad channels and applying filtering;
(2) Detecting spike events from the preprocessed recordings;
(3) Extracting waveforms around the detected spike events;
(4) Using HuiduRep to extract representations of individual spike waveforms; and 
(5) Clustering the spike representations to obtain their unit assignments.

In the pipeline, the preprocessing and threshold-based detection modules of SpikeInterface were employed to process the recordings \cite{buccino2020spikeinterface}. Following extraction, the spike representations were clustered using GMM from the scikit-learn library 
\cite{pedregosa2018scikitlearnmachinelearningpython} to produce the final sorting results.

Our pipeline is modular, meaning that each component can be replaced by alternative methods. For example, the threshold-based detection module can be substituted with more accurate detection algorithms. In the following experiments, we demonstrate that even when using a threshold-based detection module with relatively low accuracy, our pipeline still outperforms the state-of-the-art and most widely used methods such as KiloSort4.

\section{Datasets}
\subsection{International Brain Laboratory (IBL) Dataset}
The International Brain Laboratory (IBL) \cite{10.7554/eLife.63711} is a global collaboration involving multiple research institutions, aiming to uncover the neural basis of decision-making in mice through standardized behavioral and electrophysiological experiments. 

DY016 and DY009 recordings are selected from the datasets released by IBL to train and evaluate HuiduRep. Both recordings were recorded from the hippocampal CA1 region and anatomically adjacent areas. Similar to the processing in CEED \cite{NEURIPS2023_83c637c3}, we used KiloSort2.5 \cite{Pachitariu2023.01.07.523036} to preprocess the recordings and extracted a subset of spike units labeled as \textit{good} according to IBL's quality metrics \cite{banga2022spike} to construct our dataset. For every unit, we randomly selected 1,200 spikes for training and 200 spikes for evaluation. For each spike, we extracted a waveform with 121 samples across 21 channels, centered on the channel with the highest peak amplitude.

All selected units from the DY016 and DY009 recordings were used for constructing the training set.
For evaluation, we randomly sampled 10 units from the IBL evaluation dataset for each random seed ranging from 0 to 99, resulting in a total of 100 data points. 
These two subsets are referred to as the IBL train dataset and the IBL test dataset in the following sections.
\subsection{Hybrid Janelia Dataset}
HYBRID\_JANELIA is a synthetic extracellular recording dataset with ground truth spike labels designed to evaluate spike sorting algorithms. It was generated by using the KiloSort2 eMouse \cite{Pachitariu2023.01.07.523036}.
The simulation includes a sinusoidal drift pattern with $20\mu m$ amplitude and 2 cycles over 1,200 seconds, as well as waveform templates from high-resolution electrode recordings.

We evaluated model performance on both the static and drift recordings of this dataset. To ensure a fair comparison, we reported results only on spike units with the SNR greater than 3 for all models.
\subsection{Paired MEA64C Yger Dataset}
Paired\_MEA64C\_Yger is a real-world extracellular recording dataset \cite{10.7554/eLife.34518} that includes ground-truth spike times, which were obtained using juxtacellular recording \cite{PINAULT1996113}. The dataset recorded from isolated retinal tissues primarily targets retinal ganglion cells. It was collected using a 16$\times$16 microelectrode array (MEA) and an 8$\times$8 sub-array was extracted for spike sorting evaluation. For each recording, there is one ground-truth unit.

We randomly selected 9 recordings in which the ground-truth unit has SNR greater than 3, and used them to evaluate our method with other baseline models.
\section{Experiments}
\subsection{Implementation Details}
For training HuiduRep, we used the AdamW optimizer \cite{loshchilov2019decoupledweightdecayregularization} with a weight decay of $1 \times 10^{-2}$ to regularize the model and reduce overfitting.
Additionally, we employed a cosine annealing learning rate scheduler with a linear warm-up phase during the first 10 epochs, where the learning rate increased to a maximum of $1 \times 10^{-4}$.

To balance the contrastive learning branch and the DAE branch, we assigned a weight factor $\alpha$ to the denoising loss to control its contribution during training:
$$
\mathcal{L} = \alpha \cdot \mathcal{L}_{\rm{denoising}} + \mathcal{L}_{\rm{contrastive}}
$$

The model’s performance is evaluated across different values of $\alpha$ to determine the optimal trade-off on the IBL test dataset.
For each $\alpha$ setting, the learned representations were clustered using GMM, and the Adjusted Rand Index (ARI) was computed against the ground truth labels.
\begin{figure*}[t]
    \centering
    \includegraphics[width=0.7\linewidth]{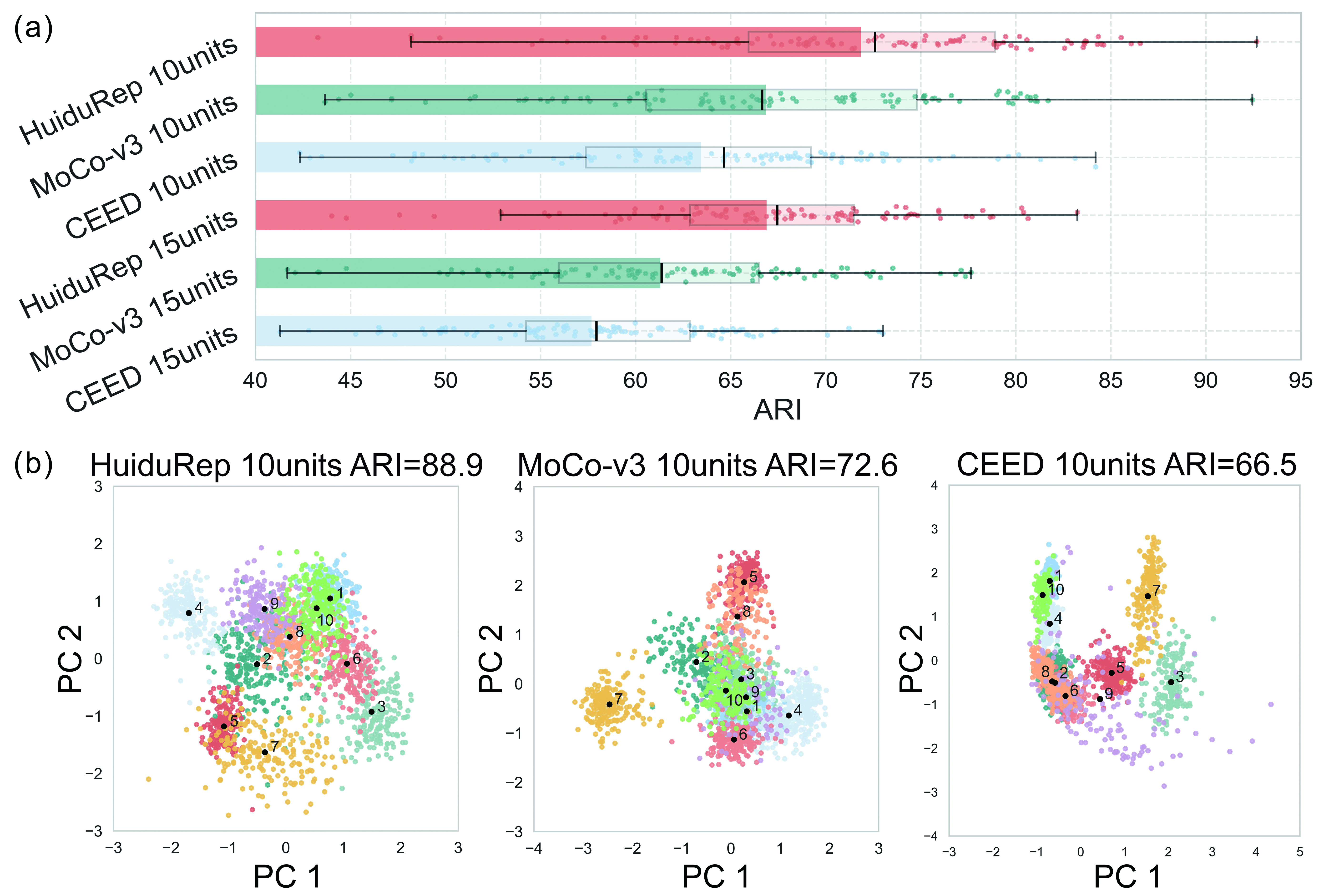}
    \caption{(a): Boxplot of HuiduRep and other models. (b): Clustering results, visualized after reduction via PCA.}
    \label{fig:ari_box}
\end{figure*}

\begin{table*}[t]
\begin{center}
    \small
    \renewcommand{\arraystretch}{1.0}
    \begin{tabular}{c|cccccc}
    \toprule
        \textbf{ARI / \ $\alpha$} & 0.0 (without Reconstruction) & 0.2 & 0.4 & 0.6 & 0.8 & 1.0 \\
    \hline
        \textbf{Mean $\pm$ SEM} & $70.5\pm1.3$ & $\mathbf{71.9\pm1.3}$ & $67.0\pm1.6$ & $71.1\pm1.4$ & $65.3\pm1.5$ & $69.6\pm1.4$\\
        \textbf{Max} & 91.5 & \textbf{92.7} & 91.0 & 91.2 & 88.8 & 90.5\\
        \textbf{Min} & 43.9 & 43.3 & 37.0 & \textbf{45.7} & 37.2 & 39.8\\
    \bottomrule
    \end{tabular}
    \caption{ARI scores (Mean $\pm$ SEM, Max, Min) across different weight factor $\alpha$ of HuiduRep, evaluated with IBL test dataset.}
    \label{tab:my_label}
\end{center}
\end{table*}
\begin{table}
    \centering
    \renewcommand{\arraystretch}{1.1}
    \setlength{\tabcolsep}{5pt}
    \small
    \begin{tabular}{c|ccc}
       \textbf{Rep Dimensions}  & 16 & 32 & 48 \\
   \hline
        \textbf{ARI} & $69.7\pm1.4$ & $71.9\pm1.3$ & $\mathbf{72.9\pm1.3}$\\
        \textbf{Time (seconds)} & \textbf{5.39 $\pm$ 0.12} & 6.78 $\pm$ 0.18 & 7.47 $\pm$ 0.23
    \end{tabular}
    \caption{ARI scores and time cost per data point (Mean $\pm$ SEM) across different representation (Rep) dimensions of HuiduRep, evaluated with IBL test dataset.}
    \label{tab:my_label}
\end{table}

We report the mean $\pm$ standard error (SEM), along with the max and min ARI values of each model across the 100 data points. The result of each data point is averaged over 50 independent GMM runs.

As shown in Table 1, the best overall performance was achieved when $\alpha = 0.2$, with the highest ARI score and the highest maximum value. Notably, both very low ($\alpha = 0.0$) and high values ($\alpha \geq 0.8$) led to decreased performance, indicating that a moderate contribution of the denoising branch is essential for improving robustness and the overall performance of HuiduRep.
\begin{table}
    \small
    \centering
    
    \renewcommand{\arraystretch}{1.0}
    \begin{tabular}{c|cc}
    \toprule
        \multicolumn{1}{c|}{Model} & ARI & Time (seconds) \\
        \hline
        HuiduRep 10units & $\mathbf{71.9\pm1.3}$ & $\mathbf{6.78\pm0.18}$ \\
        MoCo-v3 10units & $66.9\pm1.4$& $7.51\pm0.20$ \\
        CEED 10units & $63.5\pm1.3$ & $21.25\pm0.04$\\
        \hline
        HuiduRep 15units & $\mathbf{66.9\pm0.8}$ & $\mathbf{12.22\pm0.26}$\\
        MoCo-v3 15units & $61.3\pm1.1$& $13.24\pm0.28$ \\
        CEED 15units & $57.7\pm0.7$ & $24.93\pm0.09$ \\
    \bottomrule
    \end{tabular}
    \captionof{table}{ARI scores and time cost per data point (Mean $\pm$ SEM) of HuiduRep and other models across varying counts of selected units, evaluated with random seeds from 0 to 99.}
    \label{tab:ari_scores}
\end{table}
In addition, using the same IBL test dataset and evaluation method, we also evaluated the effect of different representation dimensions on the model's performance with $\alpha = 0.2$.
As shown in Table 2, with the representation dimension increasing, the model's performance generally improves, suggesting enhanced representational capacity.
However, higher-dimensional representation also leads to greater computational costs.
To balance efficiency and performance, we set the representation dimension to 32 and fixed $\alpha$ at 0.2 in all subsequent experiments.
\begin{table*}[t]
    \centering
    \renewcommand{\arraystretch}{1.1}
    \small
    \setlength{\tabcolsep}{2pt}
    \begin{tabular}{c|ccc|ccc}
    \toprule
        \multirow{2}{*}{\textbf{Method}} & \multicolumn{3}{c|}{Hybrid\_Janelia-Static (SNR $>$ 3, 9 recordings)} & \multicolumn{3}{c}{Hybrid\_Janelia-Drift (SNR $>$ 3, 9 recordings)} \\
         ~ & \textbf{Accuracy} & \textbf{Recall} & \textbf{Precision} & \textbf{Accuracy} & \textbf{Recall} & \textbf{Precision} \\
    \hline
        \textbf{HerdingSpikes2} \cite{HILGEN20172521} & $0.35\pm0.01$ & $0.44\pm0.02$ & $0.53\pm0.01$ & $0.29\pm0.01$ & $0.37\pm0.02$ & $0.48\pm0.02$\\
        \textbf{IronClust} \cite{jun_magland_2025_ironclust} & $0.57\pm0.04$ & $\mathbf{0.81\pm0.01}$ & $0.60\pm0.04$ & $0.54\pm0.03$ & $\mathbf{0.71\pm0.02}$ & $0.65\pm0.03$\\
        \textbf{JRClust} \cite{Jun101030} & $0.47\pm0.04$ & $0.63\pm0.02$ & $0.59\pm0.03$ & $0.35\pm0.03$ & $0.48\pm0.03$ & $0.57\pm0.02$\\
        \textbf{KiloSort} \cite{Pachitariu061481} & $0.60\pm0.02$ & $0.65\pm0.02$ & $0.72\pm0.02$ & $0.51\pm0.02$ & $0.62\pm0.01$ & $0.72\pm0.03$\\
        \textbf{KiloSort2} \cite{pachitariu2024KiloSort2} & $0.39\pm0.03$ & $0.37\pm0.03$ & $0.51\pm0.03$ & $0.30\pm0.02$ & $0.31\pm0.02$ & $0.57\pm0.04$\\
        \textbf{KiloSort4} \cite{pachitariu2024spike} & $0.40\pm0.03$ & $0.45\pm0.03$ & $0.52\pm0.05$ & $0.34\pm0.02$ & $0.35\pm0.02$ & $0.61\pm0.03$\\
        \textbf{MountainSort4} \cite{magland2025MountainSort4} & $0.59\pm0.02$ & $0.73\pm0.01$ & $0.74\pm0.03$ & $0.36\pm0.02$ & $0.57\pm0.02$ & $0.61\pm0.03$\\
        \textbf{MountainSort5} \cite{magland2025MountainSort5} & $0.40\pm0.06$ & $0.50\pm0.05$ & $0.52\pm0.08$ & $0.33\pm0.04$ & $0.40\pm0.03$ & $0.64\pm0.05$\\
        \textbf{SpykingCircus} \cite{10.7554/eLife.34518} & $0.57\pm0.01$ & $0.63\pm0.01$ & $0.75\pm0.03$ & $0.48\pm0.02$ & $0.55\pm0.02$ & $0.68\pm0.03$\\
        \textbf{Tridesclous} \cite{pouzat_garcia_2025_tridesclous} & $0.54\pm0.03$ & $0.66\pm0.02$ & $0.59\pm0.04$ & $0.37\pm0.02$ & $0.52\pm0.03$ & $0.55\pm0.04$\\
        \textbf{SimSort} \cite{zhang2025simsortdatadrivenframeworkspike} & $0.62\pm0.04$ & $0.68\pm0.04$ & $0.77\pm0.03$ & $0.56\pm0.03$ & $0.63\pm0.03$ & $0.69\pm0.03$\\
        \textbf{HuiduRep Pipeline without DAE} & $0.69\pm0.02$ & $0.72\pm0.02$ & $\mathbf{0.87\pm0.01}$ & $0.56\pm0.02$ & $0.61\pm0.02$ & $\mathbf{0.83\pm0.01}$\\
        \textbf{HuiduRep Pipeline with DAE} & $\mathbf{0.70\pm0.02 }^{*}$ & $0.75\pm0.02^{*}$ & $0.85\pm0.01$ & $\mathbf{0.60\pm0.02}^{*}$ & $0.65\pm0.02^{*}$ & $\mathbf{0.83\pm0.01}$\\

    \bottomrule
    \end{tabular}
    \caption{Spike sorting results (Mean $\pm$ SEM) on the HYBRID\_JANELIA dataset. Results for other methods are obtained from SpikeForest. Best-performing values are highlighted in \textit{bold}. * denote that the method performs significantly better than HuiduRep Pipeline without DAE. (Wilcoxon test, $p < 0.05$).}
    \label{tab:my_label}
\end{table*}
\begin{table*}[t]
    \centering
    \renewcommand{\arraystretch}{1.1}
    \small
    \begin{tabular}{c|ccc}
    \toprule
        \multirow{2}{*}{\textbf{Method}} & \multicolumn{3}{|c}{Paired\_MEA64C\_Yger (SNR $>$ 3, 9 recordings)} \\  
        ~ & \textbf{Accuracy} & \textbf{Recall} & \textbf{Precision}\\
        \hline
        \textbf{HerdingSpikes2} \cite{HILGEN20172521} & $0.77\pm0.10^{*}$ & $0.92\pm0.04$ & $0.80\pm0.09^{*}$ \\ 
        \textbf{IronClust} \cite{jun_magland_2025_ironclust}& $0.73\pm0.09^{*}$ & $0.96\pm0.02$ & $0.74\pm0.09^{*}$ \\
        \textbf{KiloSort} \cite{Pachitariu061481} & $0.80\pm0.09$\:  & $0.96\pm0.01$ & $\mathbf{0.82\pm0.09}$ \\
        \textbf{KiloSort2} \cite{pachitariu2024KiloSort2} & $0.69\pm0.11^{\dag}$  & $0.99\pm0.01$ & $0.70\pm0.11^{*}$ \\
        \textbf{KiloSort4} \cite{pachitariu2024spike} & $0.71\pm0.10\: $ & $\mathbf{0.99\pm0.01}$ & $0.72\pm0.11^{\dag}$ \\
        \textbf{MountainSort4} \cite{magland2025MountainSort4} & $0.80\pm0.09\: $ & $0.97\pm0.02$ & $0.81\pm0.09$ \: \\
        \textbf{MountainSort5} \cite{magland2025MountainSort5} & $0.57\pm0.10^{*}$ & $0.85\pm0.08$ & $0.60\pm0.10^{*}$ \\
        \textbf{SpykingCircus} \cite{10.7554/eLife.34518} & $0.78\pm0.10^{*}$ & $0.98\pm0.01$ & $0.79\pm0.10^{*}$ \\
        \textbf{Tridesclous} \cite{pouzat_garcia_2025_tridesclous} & $0.79\pm0.09\: $ & $0.97\pm0.02$ & $0.80\pm0.09^{*}$ \\
        \textbf{HuiduRep Pipeline with DAE} & $\mathbf{0.80\pm0.08}$ & $0.94\pm0.02$ & $\mathbf{0.82\pm0.09}$ \\
        \bottomrule
    \end{tabular}
    \caption{Spike sorting results (Mean $\pm$ SEM) on the Paired\_MEA64C\_Yger dataset. Results for other methods are obtained from SpikeForest.* and † denote that the HuiduRep Pipeline with DAE performs significantly ($p < 0.05$) and marginally significantly ($0.05 \leq q < 0.10$) better than other methods. Note: KiloSort2 was evaluated on 8 out of 9 recordings, as it is failed to run on one recording.}
    \label{tab:my_label}
\end{table*}
All models under different settings were trained for 300 epochs with a batch size of 4096 and a fixed random seed on a server with a single NVIDIA L40s 48G GPU and CUDA 12.4. A local evaluation server with a single NVIDIA RTX 5080 16G GPU and CUDA 12.8 is used to perform all experiments.
A complete list of training hyperparameters is provided in the supplementary material.

\subsection{Performance Evaluation}
To evaluate the performance of HuiduRep and other models, we created datasets where each data point includes 15 units, using the same construction method as the IBL test dataset.

As shown in Table 3, HuiduRep significantly outperforms CEED and MoCo-v3 on both the 10-unit and 15-unit test datasets, indicating superior representation learning capability. Furthermore, during testing, HuiduRep has a lower number of active parameters (0.6M) compared to CEED (1.8M). These results demonstrate that HuiduRep not only achieves better performance with reduced model complexity, but also adapts more effectively to downstream tasks such as spike sorting, which require strong representational ability.

To evaluate the performance of the HuiduRep Pipeline in real-world spike sorting tasks, two publicly available datasets, Hybrid Janelia and Paired MEA64c Yger, are selected as test sets. Multiple spike sorting tools, including KiloSort series \cite{Pachitariu2023.01.07.523036} and MountainSort series \cite{chung2017fully}, were evaluated. The performance of KiloSort4 and MountainSort5 was evaluated on our local evaluation server. The results for SimSort were cited from its original publication \cite{zhang2025simsortdatadrivenframeworkspike}, while the performance data for the remaining methods were obtained from the results provided by SpikeForest \cite{10.7554/eLife.55167}.

We recorded three metrics: accuracy, precision, and recall of different models across various test sets. Moreover, we adopted the SpikeForest definitions for computing these metrics, which slightly differ from the conventional calculation methods. The accuracy balances precision and recall, and it is similar to the F1-score. These metrics are computed based on the following quantities: $n_{1}$: The number of ground-truth events that were missed by the sorter; ${n_{2}}$: The number of ground-truth events that were correctly matched by the sorter; $n_{3}$: The number of events detected by the sorter that do not correspond to any ground-truth events. Based on these definitions, the metrics are calculated as:
\begin{align*}
\text{Precision} &= \frac{n_2}{n_2 + n_3}, \quad
\text{Recall} = \frac{n_2}{n_1 + n_2} \\
& \text{Accuracy} = \frac{n_2}{n_1 + n_2 + n_3}
\end{align*}
\begin{figure*}[t]
        \centering
        \includegraphics[width=1\linewidth]{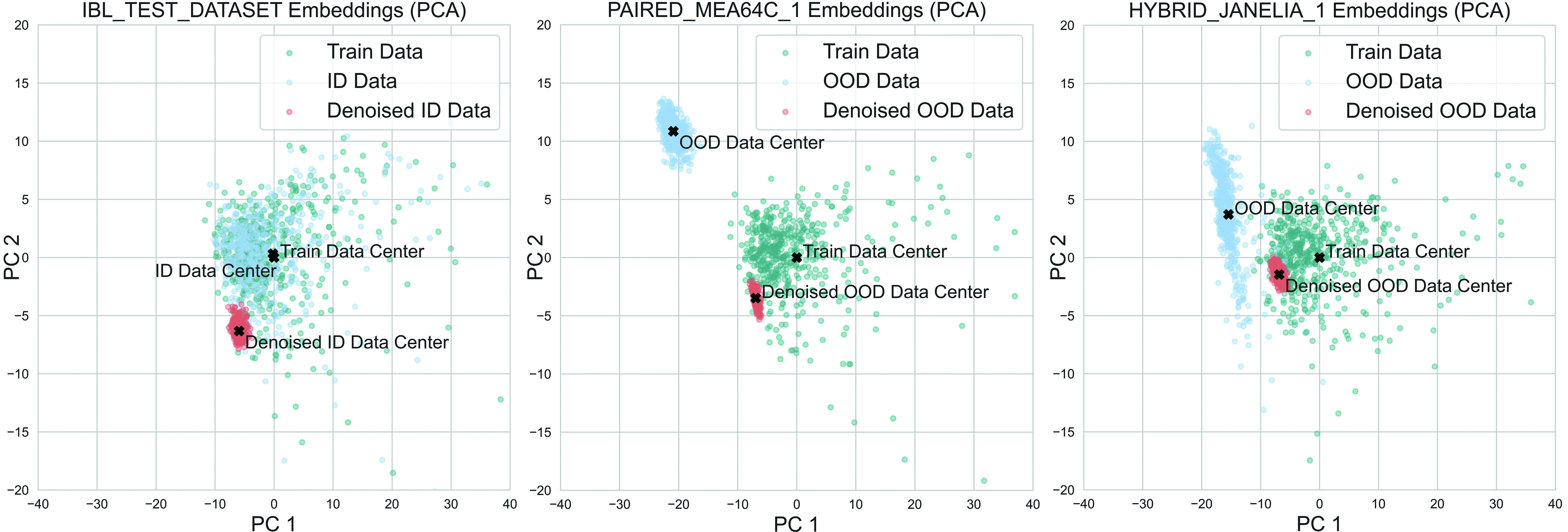}
        \caption{Reduced feature space of IBL training dataset and other datasets. The centroid of each dataset is marked with a black X. The features of each dataset are reduced to 2 dimensions using PCA.}
        \label{fig:ari_box}
\end{figure*}
\begin{table*}
    \centering
        \centering
        \small
        \renewcommand{\arraystretch}{1.1}
        \begin{tabular}{c|ccc|ccc}
        \toprule
         \multirow{2}{*}{\textbf{Dataset}} & \multicolumn{3}{|c}{without DAE} & \multicolumn{3}{|c}{with DAE}\\
         ~ & \textbf{Distance} & \textbf{Silhouette Score} & \textbf{ARI} & \textbf{Distance} & \textbf{Silhouette Score} & \textbf{ARI} \\ 
        \hline
        IBL\_Test\_Dataset & $0.46\pm0.23$ & $0.240\pm0.010$ & $0.72\pm0.03$ & $8.64\pm0.24\uparrow$ & $0.087\pm0.008\downarrow$ & $0.44\pm0.03\downarrow$\\
        Paired\_MEA64C\_Yger\_1 & $23.43\pm0.07$ & $0.176\pm0.009$ & N/A & $7.77\pm0.01\downarrow$ & $0.133\pm0.011\downarrow$ & N/A\\
        Paired\_MEA64C\_Yger\_2 & $24.72\pm0.08$ & $0.157\pm0.006$ & N/A & $7.53\pm0.02\downarrow$ & $0.120\pm0.008\downarrow$ & N/A\\
        Hybrid\_Janelia\_1 & $16.00\pm0.14$ & $0.195\pm0.011$ & $0.60\pm0.03$ & $7.02\pm0.03\downarrow$ & $0.150\pm0.005\downarrow$ & $0.64\pm0.03\uparrow$\\
        Hybrid\_Janelia\_2 & $14.53\pm0.10$ & $0.127\pm0.005$ & $0.57\pm0.02$ & $6.50\pm0.02\downarrow$ & $0.103\pm0.005\downarrow$ & $0.58\pm0.01\uparrow$\\
        Hybrid\_Janelia\_3 & $12.47\pm0.09$ & $0.159\pm0.005$ & $0.55\pm0.02$ & $6.41\pm0.02\downarrow$ & $0.131\pm0.009\downarrow$ & $0.56\pm0.03\uparrow$\\
        \bottomrule
        \end{tabular}
        \caption{Euclidean distances between the IBL training dataset and other datasets, along with silhouette score and ARI of each dataset with and without the DAE.}
        \label{tab:my_label}
\end{table*}
As shown in Tables 4 and 5, HuiduRep Pipeline consistently outperforms other models on the Hybrid Janelia dataset in terms of accuracy and precision, under both static and drift conditions. However, its recall is slightly lower than that of IronClust but significantly higher than that of the other models. This phenomenon is potentially due to threshold-based spike detection missing low-amplitude true spikes or IronClust detecting an excessive number of spikes, which leads to a high recall and lower precision.

On the high-density, multi-channel Paired MEA64C Yger dataset, the HuiduRep Pipeline also achieves slightly higher accuracy and precision compared to other models, with statistically significant or marginally significant improvements. However, the recall remains slightly lower. Detailed Wilcoxon test results are provided in supplementary material. The performance on both datasets demonstrates the practical applicability of the HuiduRep pipeline for real-world spike sorting tasks.

Notably, applying the DAE, originally an auxiliary module during training, before the contrastive learning encoder during inference leads to significant improvements in both accuracy and recall scores. We will provide an in-depth analysis of this effect in the next ablation study section.

\subsection{Ablation Study}
To investigate why the DAE enhances model performance during inference and to gain insights into its underlying mechanism, we randomly selected 500 spike samples per unit from each test dataset and the IBL training dataset. For each test dataset, the same set of samples was processed using two different methods: one with the DAE and one without. Principal Component Analysis (PCA) was then applied to reduce the dimensionality of the spike data to two dimensions. We computed the Euclidean distance between the centroid of the test samples and that of the IBL training samples in the reduced feature space. Furthermore, we applied HuiduRep followed by GMM to both groups and calculated the silhouette scores along with ARI of the resulting clusters. Since each recording in the Paired MEA64C yger dataset contains only one ground truth unit, the ARI becomes inapplicable. Each experiment was repeated 20 times, and the mean and standard deviation (STD) were reported.

As shown in Table 6 and Figure 3, applying the DAE to spike waveforms from out-of-distribution (OOD) datasets (Paired MEA64C Yger and Hybrid Janelia) significantly reduces their Euclidean distance to the IBL training set in the reduced feature space. This indicates that the DAE has learned to capture the feature distribution of the original training data. By aligning OOD data closer to the training data, the DAE effectively performs domain alignment, improving the overall ARI. Consequently, as shown in Table 3, applying the DAE before the contrastive learning encoder enables HuiduRep to better handle distribution shifts, resulting in improved accuracy and recall scores, especially on noisy and drifting recordings.

However, this benefit comes with a potential trade-off: the DAE may compress spike waveforms into a more compact space, reducing inter-class variability and thereby making them less distinguishable and slightly reducing precision scores in the subsequent spike sorting task. This effect is reflected in the decreased silhouette scores observed after applying DAE. Moreover, for in-distribution (ID) test datasets such as the IBL test dataset, the use of DAE may distort the original data distribution, resulting in increased distance to the IBL training dataset along with lower ARI. 

This suggests that while DAE effectively aligns OOD data, it may negatively impact performance when applied to data already well-aligned with the training distribution. Therefore, when processing a new dataset, one may first examine the data distribution with and without the DAE  to assess its impact on the alignment of the data.

\section{Conclusion}
In HuiduRep, the view generation strategy not only produces augmented views that preserve semantic invariance but also maintains genuine physiological significance. This strategy simulates the jitters occurring during the firing process of real neural signals, as well as the overlapping and interference between signals from different neurons. In essence, it models the natural variability present in real neural recordings. Thus, the view generation strategy encourages the model to learn spike representations under more realistic conditions, enhancing its overall performance.

Furthermore, DAE learns to reconstruct augmented inputs back onto the original spike waveforms. This component is also remarkably biologically intuitive: many cortical circuits effectively perform noise suppression and normalization. For example, computational models show that topographic recurrent networks in the cortex can amplify signal-to-noise by adjusting the excitation–inhibition balance \cite{10.7554/eLife.77009}. In other words, cortex exhibits a denoising behavior that preserves stimulus features while suppressing irrelevant fluctuations. The DAE plays a similar role: it is trained to reconstruct a clean waveform from an augmented input. In our framework, this means that HuiduRep is encouraged to represent only the stable, informative representations, effectively filtering out the noise.

Overall, HuiduRep demonstrates strong robustness in spike representation learning, outperforming state-of-the-art sorters across a wide range of datasets from distinct neural structures. By integrating contrastive learning with a denoising autoencoder, it maintains high performance under low SNR, electrode drift, and overlapping conditions. Its architecture draws inspiration from neuroscience, offering greater resilience to real-world variability than conventional methods.

While HuiduRep is designed for extracellular recordings, the core methodology, self-supervised representation learning with physiologically inspired augmentations, can generalize to other bioelectrical signals such as EMG, ECoG, and EEG. These signals share similar challenges, including low SNR, temporal variability, and inter-subject drift. Future work may explore extending HuiduRep to a broader range of electrophysiological data as well as incorporating richer biological priors and integrating more advanced signal detection techniques to further improve generalization and interpretability.

\section{Ethics Statement}
This study uses only publicly available datasets, which were collected and shared in compliance with institutional and ethical guidelines as stated in the original publications.
\section{Acknowledgments}
We acknowledge Huidu, the ragdoll cat of the co-first authors, whose name inspired the title of the HuiduRep framework.
\bibliography{aaai2026}

\end{document}